\documentclass[aps,apl,prepreint,twocolumn,superscriptaddress]{revtex4-2}

\usepackage{graphicx}
\usepackage{amsmath,amssymb}
\usepackage{hyperref}
\usepackage{physics}               
\usepackage{float}                 
\usepackage{siunitx}               
\usepackage{comment}               
\usepackage{dcolumn}               
\begin{document}

\title{Feedback-Driven Dynamical Model for Axonal Extension on Parallel Micropatterns}



\author{Kyle Cheng}
    \affiliation{Department of Physics and Astronomy, Tufts University, Medford, MA, 02155, USA}
\author{Udathari Kumarasinghe}
    \affiliation{Department of Physics and Astronomy, Tufts University, Medford, MA, 02155, USA}
\author{Cristian Staii}
    \email[Correspondence:]{cstaii01@tufts.edu}
    \affiliation{Department of Physics and Astronomy, Tufts University, Medford, MA, 02155, USA}
\begin{abstract}
Despite significant advances in understanding neuronal development, a fully quantitative framework that integrates intracellular mechanisms with environmental cues during axonal growth remains incomplete. Here, we present a unified biophysical model that captures key mechanochemical processes governing axonal extension on micropatterned substrates. In these environments, axons preferentially align with the pattern direction, form bundles, and advance at constant speed. The model integrates four core components: (i) actin–adhesion traction coupling, (ii) lateral inhibition between neighboring axons, (iii) tubulin transport from soma to growth cone, and (iv) orientation dynamics guided by substrate anisotropy. Dynamical systems analysis reveals that a saddle–node bifurcation in the actin adhesion subsystem drives a transition to a high-traction motile state, while traction feedback shifts a pitchfork bifurcation in the signaling loop, promoting symmetry breaking and robust alignment. An exact linear solution in the tubulin transport subsystem functions as a built-in speed regulator, ensuring stable elongation rates. Simulations using experimentally inferred parameters accurately reproduce elongation speed, alignment variance, and bundle spacing. The model provides explicit design rules for enhancing axonal alignment through modulation of substrate stiffness and adhesion dynamics. By identifying key control parameters, this work enables rational design of biomaterials for neural repair and engineered tissue systems.
\end{abstract}

\maketitle


\section{Introduction}

The human brain contains an estimated 100 billion neurons interconnected by axons and dendrites—structures collectively referred to as neurites \citep{Azevedo2009, Striedter2016}. Each neuron comprises a cell body, a single long axon that transmits electrical signals, and multiple dendrites that receive input from other neurons. During early development, neurites emerge from the soma, grow outward, and establish connections across the extracellular environment, ultimately giving rise to the intricate architecture of the nervous system \cite{Arimura2007, Huber2003, Lowery2009}. Among these processes, axonal growth plays a pivotal role, with axons extending over distances that span tens to hundreds of times the diameter of the cell body to reach and synapse with appropriate targets \cite{Franze2010, Kiryushko2004}. This growth is driven by the activity of the growth cone, a highly dynamic structure at the tip of the axon that interprets and responds to environmental signals \cite{Lowery2009, Goodhill2003, Franze2013}. Axonal navigation is guided by a wide array of guidance cues, ranging from diffusible molecules such as Netrins, Slit proteins, and Semaphorins to substrate-bound factors like Ephrins, extracellular matrix components, and cell adhesion molecules \cite{Franze2013,Huber2003,Moeendarbary2014, Riveline2001, Fivaz2007,Samuels1996,Toriyama2010}. These cues, along with mechanical and topographical stimuli, influence the cytoskeletal dynamics of the growth cone and steer its path through attraction or repulsion \cite{Franze2020, Alert2020}. For example, growth cones actively probe their environment for guidance cues, allowing axons to traverse long distances (often hundreds of microns) with remarkable precision, even as they encounter a dynamic and heterogeneous extracellular environment \cite{Franze2009, Franze2020, Lowery2009, Takano2019}.

\begin{figure*}
    \includegraphics[width=13cm]{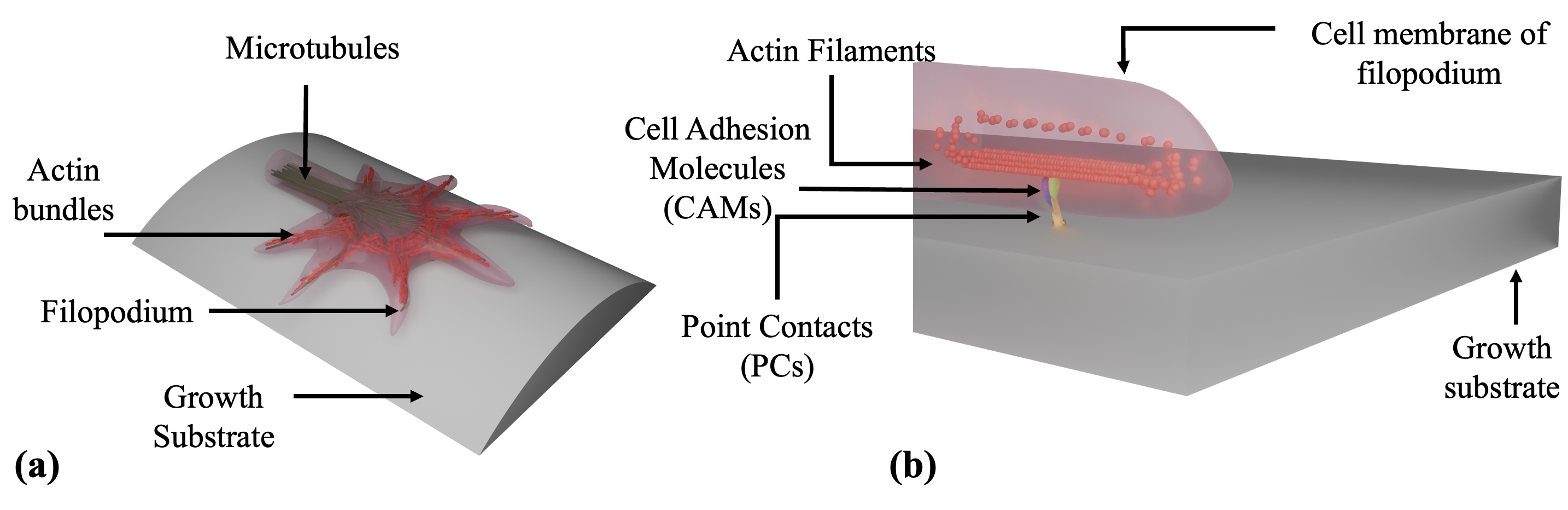}
    \caption{Schematic representation of the growth cone structure and clutch mechanism:
Panel (\textbf{a}) illustrates the major cytoskeletal components of the growth cone. Panel (\textbf{b}) depicts the molecular clutch mechanism. When the clutch is engaged, the actin cytoskeleton is mechanically linked to the growth substrate through point contacts (PCs) formed by transmembrane cell adhesion molecules (CAMs), such as integrins and cadherins. These CAMs assemble into dynamic clutch complexes that regulate retrograde actin flow and mediate adhesion to the growth substrate. The interaction among integrins, adhesion proteins, and actin filaments generates traction forces that facilitate the advancement  of the growth cone.   \label{fig1}}
\end{figure*}

\indent Robust neuronal information processing requires the ability to respond to external cues while minimizing sensitivity to random fluctuations in the surrounding microenvironment. These competing requirements are the basic characteristics of feedback control \cite{El-Samad2021, Collier1996, Takano2019}, which refers to a general class of regulatory mechanisms that biological systems employ to adapt their behavior to changing conditions. In the context of neuronal development, feedback loops are increasingly recognized as key drivers of axonal extension \cite{Oliveri2022, Descoteaux2022}. Both negative feedback, which stabilizes system output, and positive feedback, which amplifies specific signals, contribute critically to processes such as neuronal polarization and growth cone dynamics \cite{Takano2019, Fivaz2007, Samuels1996, Recho2016, Toriyama2010}. Axonal growth involves a complex interplay of biochemical signaling, cytoskeletal remodeling, and mechanical force generation, all of which are governed by tightly regulated feedback mechanisms \cite{Franze2020, Maccioni1995, Caviston2006, Coles2015, Miller1996, Lim1990, Okabe1990}. For example, growth cones convert environmental signals into directed motion by coordinating cytoskeletal dynamics and cell-substrate adhesion through a molecular clutch mechanism (Figure~\ref{fig1}) \cite{Lowery2009, Franze2010, Descoteaux2022, Medeiros2006}. In this model, actin filaments polymerize at the leading edge (actin treadmilling), while myosin II motors pull actin filaments together and induce retrograde flow \cite{Lowery2009, Franze2010, Huber2003, Medeiros2006, McKayed2013}. Transmembrane adhesion receptors such as integrins and cadherins form point contacts (PCs) with the substrate, mechanically linking the actin network to the extracellular environment and modulating traction forces \cite{Lilja2018, Polackwich2015, Pouwels2012, Jurchenko2015, Buskermolen2020, Hyland2014, Koch2012, Kumarasinghe2022}. Positive feedback arises as actin polymerization pushes the membrane forward and promotes further adhesion formation, while negative feedback emerges from the suppression of retrograde flow as adhesions mature and stabilize \cite{Lowery2009, Takano2019, Fivaz2007, Riveline2001}.

\unskip

\indent These feedback interactions allow the growth cone to dynamically sense and respond to extracellular cues, maintaining directional motion while adapting to the mechanical and geometrical properties of the substrate \cite{Buskermolen2020, Purohit2016, O'Toole2015, Lin2020, Anthonisen2019, Wang2019, Oliveri2021}. Our previous work has shown that mechanical and topographical features of the environment reinforce axonal alignment through such positive feedback mechanisms, as the growth cone continually adjusts its trajectory in response to detected substrate features \cite{Rizzo2013, Spedden2012Elasticity, Basso2019, Kumarasinghe2022,  Staii2023, Staii2024, Yurchenko2021, Yurchenko2019, Vensi2019, Descoteaux2022}. Together, these processes illustrate a delicate balance of positive and negative feedback loops that underlie robust yet adaptable axonal growth. Despite growing recognition of their importance, the precise ways in which these feedback mechanisms stabilize axonal dynamics and govern responses to fluctuating internal and external conditions remain to be fully elucidated \cite{Takano2019, Franze2020, Athamneh2015, Athamneh2017, Roy2020, Jakobs2020}.

\indent Beyond their fundamental relevance for understanding the mechanisms of axonal growth, studies of feedback regulation in neuronal development have significant implications for biomedical applications. In particular, uncovering how feedback loops coordinate cytoskeletal remodeling, traction force generation, and responses to environmental stimuli is critical for advancing nerve repair and tissue engineering strategies \cite{Pfister2011, Holland2015, Bayly2014, DeRooij2018a, DeRooij2018b, Mahar2018,  Teixeira2020}. For example, the design of neuroprosthetic devices increasingly depends on the ability to recreate growth-permissive microenvironments that simulate in vivo conditions and promote targeted axonal extension. Such efforts rely on a mechanistic understanding of how neurons interpret and respond to external cues through feedback-mediated pathways \cite{Mahar2018, Ahmadzadeh2015, Montanino2018}. These insights are also pivotal in developing innovative, bioinspired therapies for treating traumatic nerve injuries and neurodegenerative diseases \cite{Pfister2011, gladkov2017design, ishihara2011assay}. At the same time, recent advances in microfabrication and microfluidic technologies have revolutionized in vitro approaches for studying neuronal growth under controlled conditions \cite{Teixeira2020, kundu2013superimposed, gladkov2017design}. Engineered culture platforms now allow precise manipulation of biochemical, mechanical, and geometric cues, thus enabling researchers to isolate the effects of individual stimuli on axonal behavior \cite{Hyland2014, kundu2013superimposed, Jurchenko2015, ishihara2011assay, Descoteaux2022}. These tools have revealed, for instance, that substrate stiffness strongly influences axonal elongation \cite{Spedden2012Elasticity, Koch2012}, while patterned surfaces and asymmetric microchannels can guide axonal directionality and promote alignment \cite{Teixeira2020, Basso2019, Yurchenko2021, Staii2023}. Such findings are crucial for deciphering the feedback mechanisms underlying neuronal self-organization and for informing the design of biomimetic artificial neural networks \cite{Pfister2011, Mahar2018, Bayly2014, Xu2010, Staii2024}. These in vitro models serve not only as testbeds for fundamental discovery but also as platforms for engineering synthetic neural systems that replicate key functional features of the brain.

\indent In our prior work, we have systematically investigated axonal growth on poly-D-lysine-coated polydimethylsiloxane (PDMS) substrates patterned with periodic parallel ridges \cite{Basso2019, Staii2023, Staii2024, Yurchenko2021, Yurchenko2019, Vensi2019, Descoteaux2022}. These experiments demonstrated that axons preferentially align with the underlying surface features due to a deterministic torque generated by cell–substrate interactions. We further showed that axonal dynamics are modulated by feedback mechanisms, which can be altered by chemical treatments that influence adhesion and intracellular signaling pathways \cite{Descoteaux2022}. To quantitatively characterize axonal dynamics, we measured a range of mechanical and statistical parameters, including velocity and angular distributions, correlation functions, diffusion coefficients, cell–substrate interaction forces, and the axonal elastic and bending moduli \cite{Yurchenko2019, Basso2019, Yurchenko2021, Vensi2019, Descoteaux2022}. Alongside these experimental efforts, we developed a theoretical framework based on Langevin and Fokker–Planck equations to model growth cone dynamics and predict how external signals influence neuronal behavior \cite{Rizzo2013, Descoteaux2022, Staii2023, Staii2024}. Our findings indicated that axonal growth on flat poly-D-lysine-coated glass surfaces can be described by linear Langevin equations with stochastic noise, leading to a collective regulatory behavior of axonal speeds. We also applied these models to ratchet-like substrates composed of asymmetrically tilted nanorods, allowing us to calculate effective diffusion coefficients and analyze directional biases in growth \cite{beighley2012alignment, Spedden2014Asymmetry}. In these studies, we have emphasized that axonal guidance on micropatterned surfaces arises from the interplay between stochastic fluctuations and deterministic cues such as substrate geometry and mechanical constraints. These insights support the view that neuronal growth is governed by dynamic feedback loops, wherein growth cones continuously sense and respond to their microenvironment, adjusting their trajectory in real time.

\indent In this paper, we present a biophysical model of axonal dynamics that incorporates actin and tubulin transport, cell–substrate adhesion forces, and lateral inhibition between neighboring axons. Unlike earlier models, which were largely phenomenological, our approach couples three interacting dynamical subsystems: (1) an actin–adhesion feedback loop that generates a traction switch, (2) Semaphorin–Slit–mediated lateral inhibition between axons, and (3) soma-to-growth-cone tubulin transport that determines the rate of axonal elongation. In addition, we introduce cell-substrate mechanical coupling that converts traction forces into an alignment torque. The resulting six-variable model, supplemented with orientation noise, is analyzed using bifurcation theory, invariant manifolds, and stochastic averaging. Our analysis reveals that (i) a saddle-node bifurcation generates a high-traction branch, (ii) traction feedback shifts a pitchfork bifurcation to select between alternating high- and low-repulsion stripes, (iii) the tubulin subsystem admits a family of exact linear solutions that become globally attracting, and (iv) axons align with the micropatterned grooves while exhibiting a quantifiable diffusion coefficient. Finally, we demonstrate that the model's predictions are consistent with experimental observations from traction force microscopy and live-cell imaging, showing that axons self-organize and extend preferentially along the micropatterns at an approximately constant speed. 
\\

\section{Materials and Methods}

\textit{Neuronal Cell Culture}. Primary cortical neurons were obtained from embryonic day 18 rat embryos. All procedures involving animal tissue were approved by the Tufts University Institutional Animal Care and Use Committee and were conducted in accordance with NIH guidelines for the Care and Use of Laboratory Animals. Neuronal dissociation and culture were performed using established protocols described in our previous publications \cite{Rizzo2013, Spedden2012Elasticity, Basso2019, Kumarasinghe2022,  Staii2023, Staii2024, Yurchenko2021, Yurchenko2019, Vensi2019, Descoteaux2022}. Immunostaining data from earlier studies confirmed high neuronal purity in these cultures \cite{Spedden2012Elasticity}. Cells were plated onto micropatterned polydimethylsiloxane (PDMS) substrates pre-coated with poly-D-lysine (PDL; 0.1 mg/mL, Sigma-Aldrich, St. Louis, MO) at a density of 4,000 cells/cm². As demonstrated in previous work, neuronal cultures maintained at low densities (3,000–7,000 cells/cm²) promote the development of long axons, making them well-suited for studying axonal dynamics under controlled surface cues \cite{Basso2019, Kumarasinghe2022,  Staii2023, Staii2024, Yurchenko2021, Yurchenko2019, Vensi2019, Descoteaux2022}.

\begin{figure*}
    \includegraphics[width=10 cm]{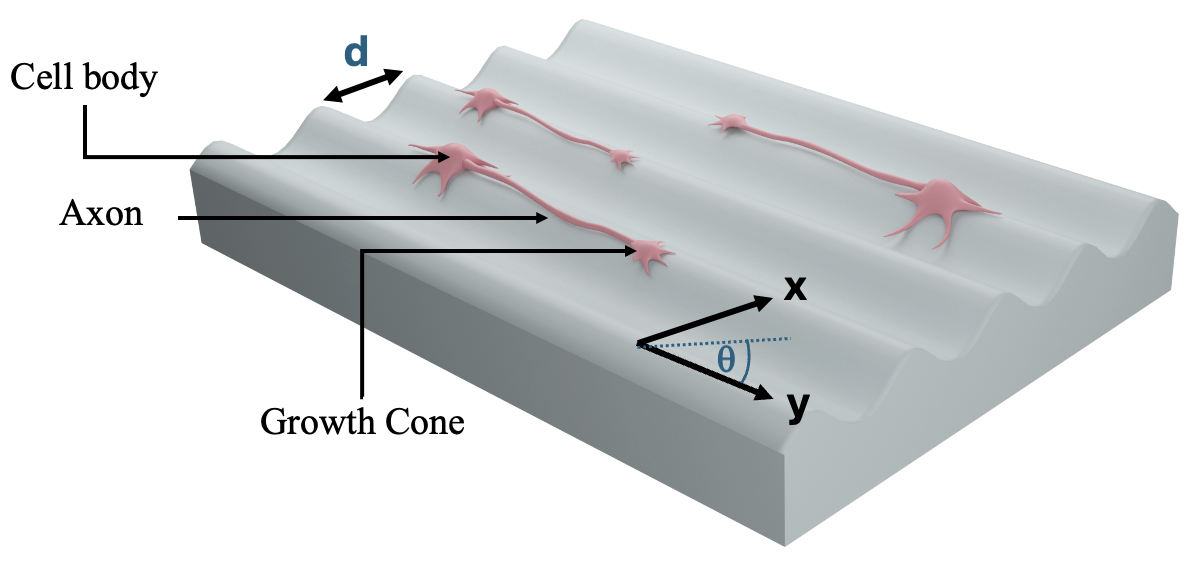}
    \caption{Schematic of a micropatterned PDMS surface. The image shows periodic surface patterns in the $x$ direction, with spatial period $d$ and uniform peak height. The $y$-axis is defined as parallel to the direction of the PDMS grooves. Cortical neurons are shown on top of the micropatterns, extending a long axon aligned with the grooves and several shorter dendrites. Axonal growth is directed by the growth cone. The growth angle, $\theta(t)$, is defined as the angle between the axonal velocity vector and the $y$-axis at time t. \label{fig2}}
\end{figure*}
\unskip
\hfill \break

\textit{Micropatterned Substrates}. Micropatterns on the PDMS substrates consisted of parallel ridges separated by grooves (Figure~\ref{fig2}). These patterns were fabricated using a simple imprinting method in which diffraction gratings were pressed into uncured PDMS, producing periodic structures with a defined spatial period $d$. The PDL coating was applied via spin coating to ensure uniform surface treatment. Further details on substrate preparation and micropatterning techniques are provided in our previous publications \cite{Basso2019, Kumarasinghe2022,  Staii2023, Staii2024, Yurchenko2021, Yurchenko2019, Vensi2019, Descoteaux2022}.

\textit{Imaging and Data Acquisition}. Atomic force microscopy (AFM) and fluorescence imaging were used to characterize both the substrates and neuronal growth. AFM images were acquired using an MFP-3D system (Asylum Research) equipped with a BioHeater fluid cell, integrated with an inverted Nikon Eclipse Ti microscope (Micro Video Instruments, Avon, MA). Fluorescence imaging of neurons was performed using standard FITC filters (excitation: 495 nm; emission: 521 nm).

\textit{Data Analysis}. Growth cone dynamics were analyzed using ImageJ (NIH). The position of each growth cone was tracked by fluorescence microscopy, with images captured every $\Delta t = 5$ minutes over a total observation period of 30 minutes. Measurements were performed at multiple time points post-plating: t = 10, 15, 20, 25, 30, 35, 40, 45, and 50 hours. The chosen interval ($\Delta t = 5$ minutes) ensured that the displacement $\Delta r$ of the growth cone exceeded the spatial resolution of the measurement (~0.1 µm), and that the velocity estimate $\Delta r/\Delta t$  closely approximated the instantaneous growth velocity $V$. The growth angle $\theta$ was defined relative to the $y$-axis, as illustrated in Figure~\ref{fig1}.

\textit{Numerical simulations}. We simulate axon trajectories using the stochastic Euler--Maruyama method \cite{pearson2011modeling, amselem2012stochastic, li2011dicty, Staii2024}. The model consists of $N = 10$ parallel micropatterns, each containing a single growth cone. Growth cone $i$ interacts with its immediate neighbors ($i-1$ and $i+1$) as well as with its underlying micropattern. The position of each growth cone is described by the arclength $s$ measured from the axon's origin. At each time step, the turning angle, which represents the axon's steering behavior, is determined by the local interactions and a stochastic term modeled as an uncorrelated Wiener process. The growth velocity is computed from the displacement of the growth cone at each step.

\section{Results}
\begin{figure*}
    \includegraphics[width=12 cm]{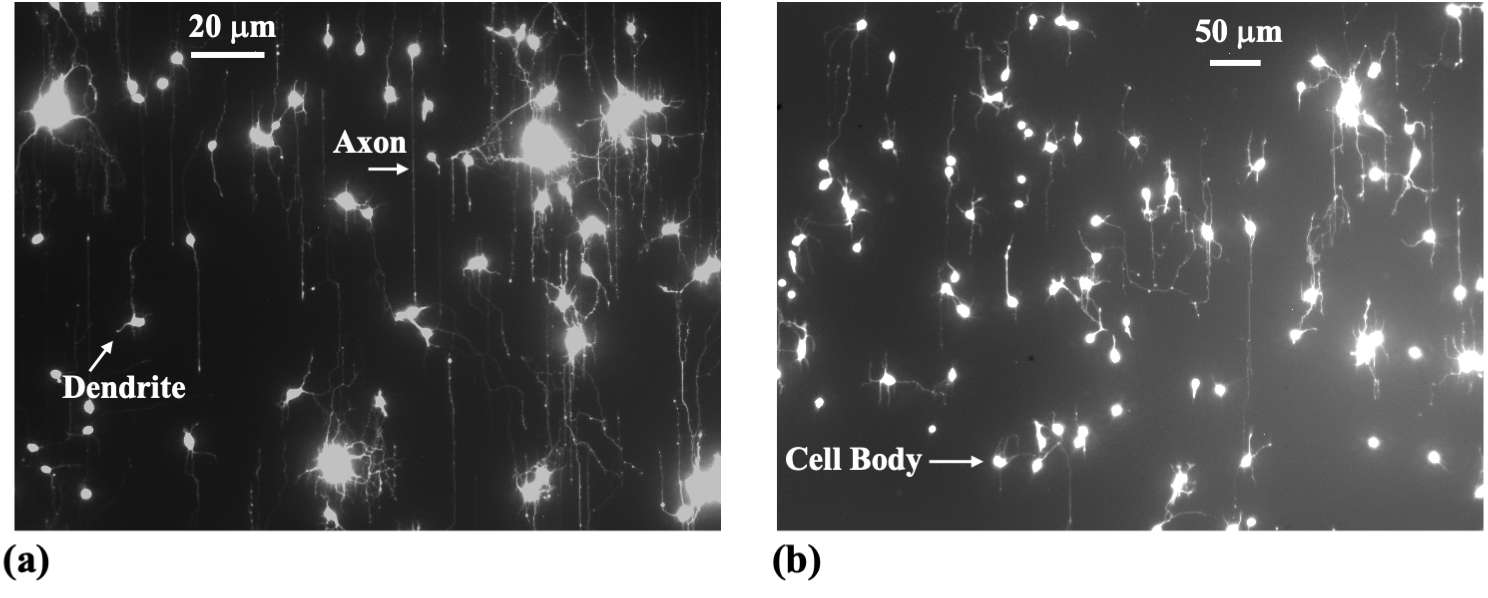}
    \caption{Fluorescence images (Tubulin Tracker Green) illustrating axonal growth in cortical neurons cultured on poly-D-lysine (PDL)-coated PDMS substrates with periodic micropatterns, with spatial period $d=3 \mu m$ (in \textbf{a}) and $d=5 \mu m$ (in \textbf{b}). The images reveal strong axonal alignment along the direction of the micropatterns \cite{Yurchenko2021, Staii2023}. Also shown in the figure are the major structural elements of a neuronal cell: cell body, axons, and dendrites.  \label{fig3}}
\end{figure*}
\unskip

\subsection*{\textit{3.1 Model parameters.}}
Axonal growth on engineered micropatterns provides a well-controlled experimental platform for probing the cell–cell signaling processes that shape neuronal network formation. Figure~\ref{fig3} shows examples of fluorescent images for axonal growth on PDMS substrates with  $d=3 \mu m$ (Figure~\ref{fig3}a) and $d=5 \mu m$ (Figure~\ref{fig3}b). Complementing these experiments, mathematical models offer a powerful framework for testing hypotheses and elucidating the biophysical principles underlying neuron development \cite{Descoteaux2022, Staii2024, Samuels1996, Oliveri2022, DeGennes2007, Murray1993}. More broadly, neurite growth results from a complex interplay of coupled mechanisms, including elasticity, cellular adhesion, traction force generation, intracellular transport, and chemical signaling, that remain only partially understood. This complexity motivates the modeling of growth in terms of the active transport of key cytoskeletal components (tubulin and actin), the density of point contacts (PCs) and cell adhesion molecules (CAMs), mechanical strain, and the generation of traction forces. In this study, we present a mathematical model of axonal dynamics that incorporates the following key biophysical parameters governing neuronal growth on micropatterned substrates:\\

\noindent
\textbf{Biochemical–mechanical state of a single growth cone:}\\
$A(t)$: polymerized F-actin density;\\
$C(t)$: density of active CAM-PC complexes (effective adhesion sites); \\
$F(t)$ traction force transmitted to the substrate.
\\

\noindent
\textbf{Growth cone repulsion (lateral inhibition}):\\
$I_i(t)$: repulsive signal (Netrin, Ephrin) \textit{received} by a growth cone on micropattern $i$;\\
$S_i(t)$: repulsive ligand (Semaphorin3A or Slit proteins) \textit{emitted} by a growth cone on micropattern $i$.
\\

\noindent
\textbf{Tubulin transport model:}\\
$c_0(t)$: tubulin concentration in the soma;\\
$c_1(t)$: tubulin concentration in the growth cone;\\
$\ell(t)$: axonal length.
\\

The principal parameters are listed in Table \ref{tab:param}. 

\begin{table*}
\centering
\begin{tabular}{@{}ll@{}}
\hline
Symbol & Meaning \\ \hline
$\alpha,\beta$ & Actin and adhesion turnover ratios \\
$k$            & Force constant in $F=kAC$ \\
$a$            & Ratio between pattern spatial period $d$ and growth cone dimension $l_0$ \\
$\nu$          & Ligand/receiver lifetime ratio \\
$\gamma$       & Coupling strength between traction and ligand export \\
$\lambda,\varepsilon$ & Orientation-torque coefficient and stiffness anisotropy \\
$\alpha_t,\beta_t,\gamma_t$ & Parameters of the tubulin model (see governing equations) \\
$D_\theta$       & Angular diffusion coefficient (orientation-noise strength) \\ \hline
\end{tabular}
\caption{Main model parameters. }
\label{tab:param}
\end{table*}

In this model, all variables are rescaled to remove units, making the parameters above dimensionless representations of concentrations, coupling strengths, and lengths. 
\\

\subsection*{\textit{3.2 Governing equations.}} 
\noindent
\textbf{Generation of traction forces.} Growth cones translate extracellular cues into directed motion by
coordinating cytoskeletal dynamics and cell–substrate adhesion through a molecular clutch mechanism.  Moreover, Traction Force Microscopy (TFM) experiments indicate that the contractile force
transmitted to the substrate scales with both actin and PC-CAM density \cite{Hyland2014, Koch2012, Kumarasinghe2022}. Based on this framework we introduce the following mathematical model for actin--adhesion generated traction:

\begin{align}
\dot{A} &= -\alpha A + C, \tag{1}\\
\dot{C} &= \frac{A^{2}}{1+A^{2}} - \beta C, \tag{2}\\
F &= kAC \tag{3}
\end{align}

 In this model, actin depolymerizes with dimensionless rate $\alpha$ and is promoted by nascent
adhesions, whereas adhesion assembly is a sigmoidal function of actin and
turns over at rate $\beta$ \cite{Huber2003, Lowery2009, Franze2010, Kiryushko2004}. The parameter $k$ converts contractile stress into traction force $F$. In addition, since the micropatterns present a higher effective stiffness parallel to their
axis (direction $y$ in (Figure~\ref{fig2}), the growth angle $\theta$ relative to the micropattern satisfies the following equation \cite{Descoteaux2022}:

\begin{equation}
\dot{\theta} = -\lambda(1-\varepsilon)F\sin (\theta) + \eta(t),\qquad
\langle \eta(t)\eta(t') \rangle = 2D_{\theta}\,\delta(t-t'),\tag{4}
\end{equation}
with $\lambda$ a torque coefficient, $\varepsilon$ the stiffness anisotropy, and $D_{\theta}$ the angular diffusion coefficient. The stochastic term $\eta$ is represented by Gaussian white noise with zero mean \cite{Staii2023, Staii2024}.

Equations~(1)--(4) are a direct dynamical
translation of the clutch hypothesis: actin treadmilling adds protrusive
filaments ($A$), newly engaged clutches ($C$) supply resistance necessary for
force build‑up, and the product $A\,C$ quantifies the number of
force‑bearing links that together generate substrate traction $F$. The
negative term $-\alpha A$ captures actin retrograde flow, while the saturation in
Equation ~(2) reflects cooperative clutch formation whose rate
plateaus once actin is accumulated at saturating levels.  In short, the subsystem encodes the
positive feedback: ``more actin $\Rightarrow$ more clutches
$\Rightarrow$ stronger resistance $\Rightarrow$ more actin'' that powers
axon advance, balanced by turnover rates $\alpha$ and $\beta$.\\

\noindent
\textbf{Growth cone lateral inhibition.}
To model the near-neighbor axonal inhibition we adapt the "Delta-Notch" lateral-inhibition model of Collier \textit{et al.} \cite{Collier1996}, and write the following coupled equations for received inhibition $I_i$ and emitted inhibitory ligand $S_i$ for a growth cone on micropattern $i$ ($1\leq i \leq N$, where $N$ is the total number of micropatterns):
\begin{align}
\dot{I}_i &= f(\bar S_i) - I_i, \tag{5}\\
\dot{S}_i &= \nu\bigl[g(I_i) - S_i\bigr]e^{-\gamma F_i}, \tag{6} \\
f(x)&= \frac{x^{2}}{a+x^{2}}, \ g(x)=\frac{1}{1+a\,x^{2}} \tag{7}\\
F_i &= kA_iC_i.\tag{8}
\end{align}
In these equations, $f$ is a strictly increasing function, $g$ is strictly decreasing,
and $\nu=r/m$ is the ratio of decay rates for $S$ and $I$ \cite{Collier1996}. The local average $\bar S_i = \tfrac12\bigl(S_{i-1} + S_{i+1}\bigr)$ appears in Equation ~(5) because each growth cone senses the inhibitory molecules predominantly through filopodia found in the proximity of its immediate left "$i-1$" and right "$i+1$" neighbors. The mean therefore captures the first--order external signal that drives inhibition activation. "High--S" cones advance along the micropattern, while "High--I" ("Low--S") cones pause and explore alternative trajectories. We couple this model to the mechanical Equations ~(1)--(3) by using the Hill functions $f(x)$ and 
  $g(x)$ (Equations~(7)). In our previous work \cite{Vensi2019}, we experimentally demonstrated that axons exhibit maximum degree of directional alignment on micropatterned surfaces when the surface spatial period $d$ (Figure~\ref{fig2}) matches the characteristic size of the growth cone $l_0$, corresponding to $a\approx 1$ in our model.  We assume that high-traction growth cones emit fewer repulsive ligands, thereby linking the mechanical model to the inhibitory signaling pathway through the dimensionless coupling parameter $\gamma$.  \\   

\noindent
\textbf{Tubulin transport.}
A key factor in neurite growth is the availability of tubulin, which polymerizes to form microtubules that support axonal extension. During the early stages of axonal development, tubulin and other essential cellular components are synthesized in the soma and must be delivered to the growth cone. This transport occurs through a combination of diffusion and active motor-driven mechanisms along the axonal shaft \cite{Huber2003, Lowery2009, Franze2010, Kiryushko2004}. A common modeling approach represents the neuron as a small number of compartments, each characterized by specific chemical concentrations \cite{Samuels1996, Oliveri2022}. These compartments exchange materials via diffusion and active transport driven by molecular motors. Following the model introduced by Oliveri \& Goriely~ \cite{Oliveri2022}, we adopt a simplified two-compartment framework in which the compartments are separated by a distance 
$l$, representing the axonal length. In \textit{dimensionless} form, the resulting coupled differential equations describe tubulin transport from the soma to the growth cone as follows \cite{Oliveri2022}:

\begin{align}
\dot{c}_0 &= 1 - \alpha_t\frac{c_1 - c_0}{\ell}, \tag{9}\\
\dot{c}_1 &= -\gamma_t c_1 + \beta_t + \alpha_t\frac{c_1 - c_0}{\ell}, \tag{10}\\
\dot{\ell} &= \gamma_t c_1 - \beta_t,\tag{11}
\end{align}

\noindent
where $c_0$ and $c_1$ are the (scaled) tubulin concentrations in the soma and growth cone, respectively, and $\ell$ is the axonal length. The parameter $\alpha_t$ represents the tubulin transport rate, $\gamma_t$ the local consumption rate, and $\beta_t$ the treadmilling offset. A detailed derivation, along with a discussion of the key properties of these equations, is provided in the Appendix B. \\

\subsection*{\textit{3.3 Dynamical--systems analysis}}

The system of coupled nonlinear differential Equations ~(1)-(11) represents a complete dynamical model that integrates the key mechanical, biophysical, and biochemical parameters that govern axonal growth on the micropatterned substrates. In this section, we apply the theory of dynamical systems to analyze this model. Specifically, we examine the global phase portrait, identify exact linear solutions, characterize stable and unstable saddle nodes, locate bifurcation points, and outline the conditions that determine whether the system undergoes transient collapse or sustained axonal elongation. Additional details of the analysis and derivations are provided in the Appendix A and Appendix B. 

Figures~\ref{fig4}a and ~\ref{fig4}b show the solutions of Equations~ (5)-(7) for the case where the dimensionless parameter $a\approx 1$, which corresponds to the micropattern spatial period $d$ matching the characteristic size of the growth cone $l_0$. Under these conditions, the growth cone dynamics stabilize at high--S (low--I) values of the inhibitory signals, indicating robust advancement of the growth cone along the micropattern, consistent with the experimental observations \cite{Vensi2019, Yurchenko2019, Yurchenko2021}. Furthermore, the tubulin transport model shows that the sign of the elongation rate $\dot{\ell}$ is determined by the difference between the tubulin concentration in the growth cone $c_1$ and the threshold parameter $\Lambda = \beta_t / \gamma_t$ (Equation~(10)). Specifically, when the initial tubulin concentration satisfies $c_1(0)< \Lambda$, the axon initially undergoes collapse until $c_1$ increases sufficiently to reach the threshold, at which point steady axonal growth begins.

Figures~\ref{fig4}c and \ref{fig4}d illustrate these dynamics. The green curve in Figure~\ref{fig4}c shows axonal length $\ell$  for $\Lambda=2$. In this case, the initial tubulin concentration in the growth cone is below the threshold $c_1(t)<\Lambda$, causing the axon to retract initially. As retrograde transport and local tubulin concentration $c_1$ increase, it eventually exceeds the threshold, at which point $\dot\ell$ becomes positive and linear elongation begins. The blue curve in Figure~\ref{fig4}c corresponds to $\Lambda=0.25$, when the threshold is already met at $t=0$, so no collapse occurs and the axon grows continuously. The red curve, representing the intermediate case $\Lambda=1$, shows a borderline scenario with a brief plateau before growth resumes.
Figure~\ref{fig4}d shows the corresponding tubulin concentration $c_1$ in the growth--cone for each case, confirming the described behavior. Lower values of $\beta_t$ (and thus smaller $\Lambda$) result in slower delivery of tubulin from the soma to the growth cone, delaying the point at which
$c_1$ surpasses the critical threshold. Despite these initial differences, all solutions ultimately converge to the same steady-state behavior, consistent with the experimentally observed constant elongation speed at large axonal lengths.  \cite{Vensi2019, Yurchenko2019, Yurchenko2021}.

\begin{figure*}
    \includegraphics[width=12 cm]{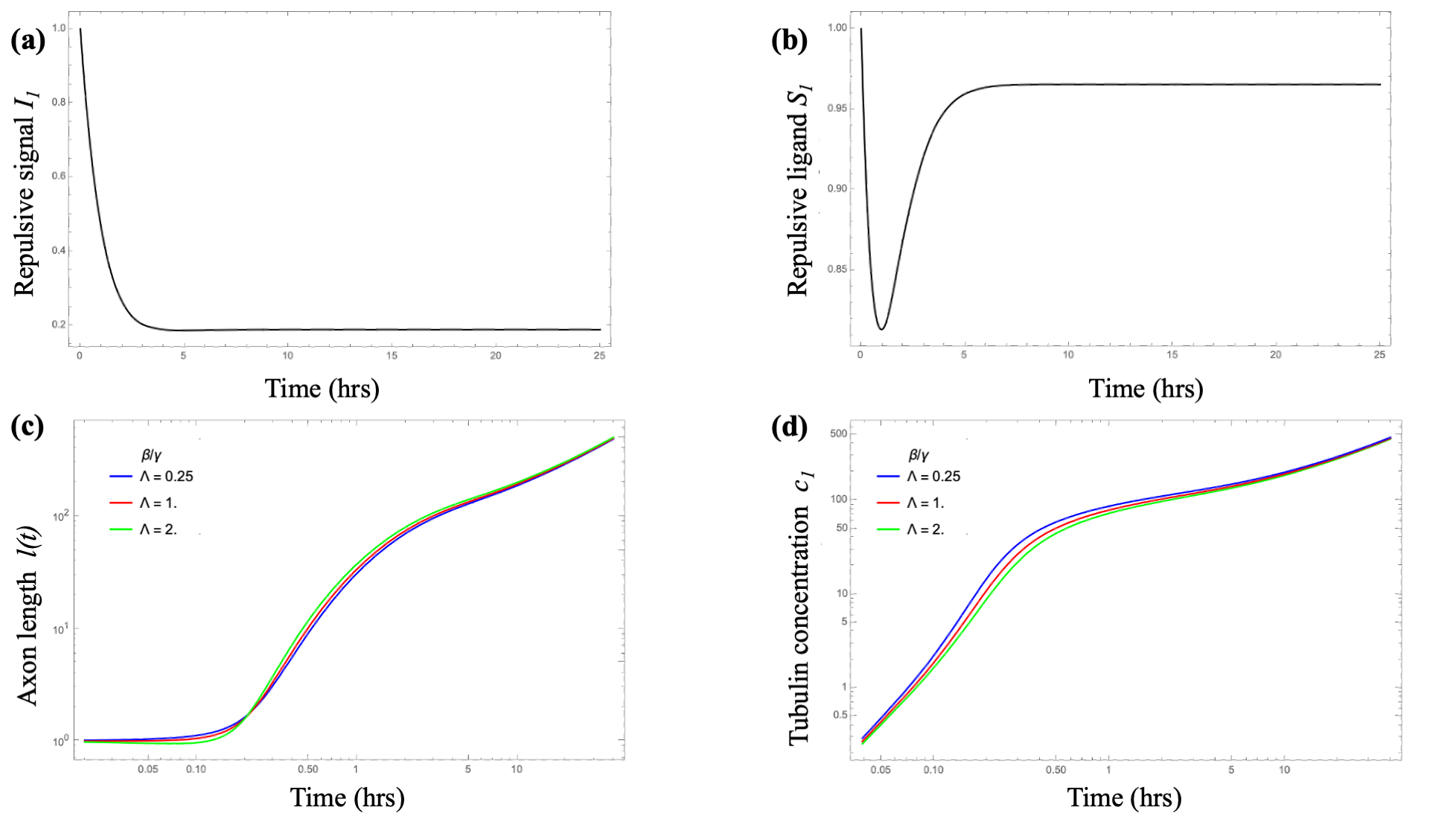}
    \caption{(\textbf{a,b}) Solutions of the lateral inhibition model for the case where the dimensionless parameter $a \approx 1$, corresponding to the micropattern spatial period $d$ matching the characteristic growth cone size $l_0$. The plots show that the system stabilizes at high–S (low–I) levels of inhibitory signaling. This regime supports sustained growth cone advancement along the micropattern, in agreement with experimental observations. \textbf{(c)} Log--log plot of axon length $\ell(t)$ for three different values of the parameter $\Lambda = \beta_t / \gamma_t$: $\Lambda=0.25$ (blue), $\Lambda=1$ (red) and $\Lambda=2$ (green). For $\Lambda=2$ the curve initially moves downward, indicating a transient collapse, then turns upward once $c_1$ exceeds the threshold. The marginal case
  $\Lambda=1$ shows an almost flat segment before eventual growth,
  whereas $\Lambda=0.25$ grows monotonically from the outset. (\textbf{d}) Time evolution of the tubulin concentration in the growth cone $c_1(t)$ for the same
  parameter sets and colors as in \textbf{(a)}.
  Larger $\Lambda$ values draw more tubulin to the growth cone,
  producing an increase in $c_1$ that parallels the  growth phase in the
  axonal length plot. Eventually all three trajectories converge toward the common
  steady state dictated by the transport model Equations~(9)-(11). \label{fig4}}
\end{figure*}   
\unskip

The steady states of the biomechanical Equations (1)-(3) satisfies: $C = \alpha A$ and $A = (1 + A^{2})\alpha\beta$. 
The discriminant of this quadratic equation vanishes at $\alpha\beta = \tfrac12$, producing stable high--traction nodes $(A_+,C_+)$ and unstable saddle $(A_-,C_-)$ (the linear stability of this system is analyzed in Appendix A). The stable nodes represent a persistently motile growth cone, while the saddle point acts as a separatrix: once the trajectory crosses this boundary, the growth cone transitions into a state of sustained advancement along the micropattern. This actin-mediated saddle–node structure partitions the phase space into two distinct regions: a “protrusive” and a “non-protrusive” domain. Crossing the separatrix triggers a sharp onset of traction, signaling the transition to directed growth.
 
\begin{figure*}
    \includegraphics[width=12 cm]{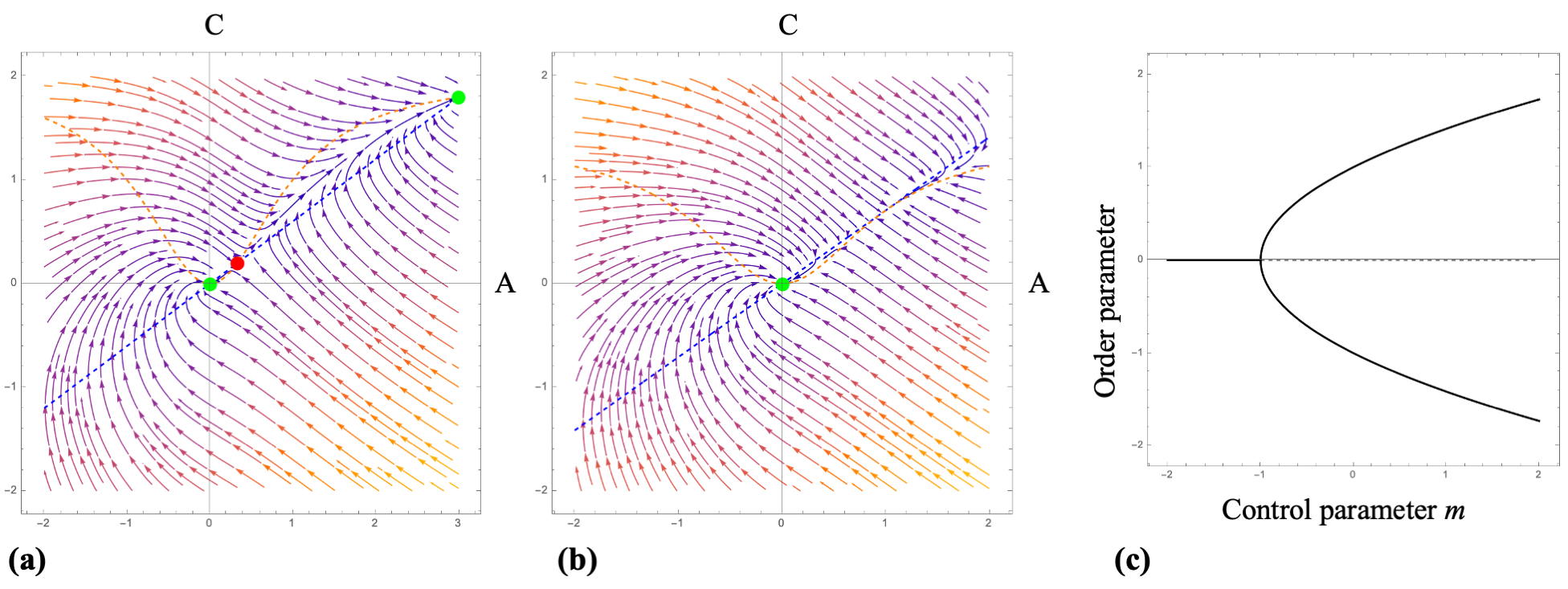}
    \caption{\textbf{(a)} Phase portrait of the system with $\alpha=0.6$, $\beta=0.5$.  The arrows represent the vector field $\bigl(\dot A,\dot C\bigr)$. The dashed curves are the nullclines: $\dot A=0$ (blue) and $\dot C=0$ (orange). The green dots represent the two stable nodes $E_0=(0,0)$ and $E_1=(3,\tfrac95)$. The red dot is the saddle point $E_s=(\tfrac13,\tfrac15)$. \textbf{(b)} Phase portrait of the system at the critical turnover product $\alpha\beta=1/2$, which illustrates the loss of bistability. The two non-trivial equilibrium points shown in (\textbf{a}) coalesce, and all the trajectories flow towards the quiescent state at the origin (green dot). \textbf{(c)} Bifurcation diagram, showing the emergence of two stable equilibrium branches when the control parameter exceeds the critical threshold value $m>-1$. Solid black branches denote stable equilibria. Dashed branches denote unstable equilibria. The single stable state corresponding to homogeneous signaling becomes unstable when the condition $m= -1$ is satisfied. Beyond this threshold, two solutions appear smoothly, without any discontinuous change in the order pattern, consistent with a supercritical pitchfork bifurcation. \label{fig5}}
\end{figure*}

Figure~\ref{fig5}a shows the two-dimensional phase portrait in the $(A, C)$ plane for parameter values $\alpha = 0.6$ and $\beta = 0.5$, corresponding to the regime $\alpha \beta < 1/2$. The arrows represent the vector field $\bigl(\dot{A}, \dot{C}\bigr)$, indicating the direction and magnitude of the system's dynamics at each point in the plane. Trajectories are tangent to these arrows, providing a complete depiction of the phase flow under the specified parameters. The nullclines of the system are also shown: the blue dashed curve corresponds to $\dot{A} = 0$ (actin turnover balance), and the orange dashed curve to $\dot{C} = 0$ (adhesion turnover balance). Equilibria occur at the intersections of these curves, where both derivatives vanish.

\unskip

Solving $\dot{A} = \dot{C} = 0$ yields three equilibrium points: two stable nodes, $E_0$ and $E_1$ (green dots), and a saddle point, $E_s$ (red dot). The flow lines that converge toward/diverge from $E_s$ form the \textit{separatrix}, which divides the plane into two basins of attraction (Figure~\ref{fig5}a). Initial conditions on the left or below the separatrix evolve toward $E_0$, a collapsed state with low actin and weak adhesions. In contrast, initial states on the right or above the separatrix converge to $E_1$, characterized by high actin levels and strong adhesions, representing a protrusive, strongly adhered state. Increasing either $\alpha$ or $\beta$ can eliminate $E_1$ through a saddle-node bifurcation, leaving only the collapsed equilibrium.

Figure~\ref{fig5}b illustrates the phase portrait at the critical value $\alpha \beta = 1/2$, where the two non-trivial equilibria coalesce. This bifurcation results in the loss of bistability: all trajectories now converge to the origin $E_0$, and the system no longer supports the protrusive state.

Linearizing Equations~(5)-(8) and Fourier transforming $\{I_i,S_i\}\propto e^{ikx}$ yields a pitchfork bifurcation threshold with an effective control parameter (Appendix A):

\begin{equation}
m \;=\; -\,f'(I_0)\,g'(S_0)\,e^{-\gamma F} \tag{12}
\end{equation}

Bifurcations typically signify qualitative changes in a system's dynamics as control parameters are varied \cite{Murray1993, strogatz2015nonlinear}. Among them, pitchfork bifurcations are characteristic of systems with underlying symmetry, such as the present case involving parallel micropatterns with a constant spatial period $d$. In a pitchfork bifurcation, a symmetric fixed point loses stability and gives rise to two new, symmetry-related equilibrium states. The bifurcation is classified as \emph{supercritical} if these new branches are stable and emerge continuously beyond the critical parameter value \cite{Murray1993, strogatz2015nonlinear}.  

Our model demonstrates that for $m<-1$, the system exhibits a stable uniform state characterized by the order parameter \(y=0\). In this regime, trajectories converge to this homogeneous solution. At the critical value \(m=-1\) this fixed point loses stability, and two new equilibria emerge symmetrically at \(y=0^{\pm}\) (Figure~\ref{fig5}c). When \(m>-1\), the point \(y=0\) becomes unstable, while the two stable points move along \(y=\pm\sqrt{m+1}\), drawing nearby axonal trajectories. The smooth emergence and stability of these branches indicate a super‑critical pitchfork bifurcation, supporting a pattern‑formation mechanism that enforces the mechanical alignment imposed by the micropatterns.\\

\subsection*{\textit{3.4 Connecting the dynamical-systems analysis to axonal growth.}}

From Equations (9)-(11) we obtain that the combination $\mathcal J(t)=c_0(t)+c_1(t)+\ell(t)-t$ obeys $\dot J = 0$, such that $J = J_0$ is an affine invariant conserved along every trajectory. Choosing $J_0=1$ and solving the remaining linear differential equation gives the exact solution (Appendix B):
\begin{align}
(c_0,c_1,\ell) &= \bigl(c_0, \; c_{00} + c_{01}t,\; vt\bigr) \nonumber \\ 
v &= \frac{\sqrt{\Lambda(\Lambda + 4)} - \Lambda}{2},\  \nonumber \\ 
\Lambda &= \beta_t/\gamma_t \tag{13}
\end{align}

As discussed in the previous section, linearization shows that this branch is globally attracting: irrespective of initial conditions, $\ell$ increases linearly at speed $v$ after a transient.

Writing the deterministic part in Equation~(4) near a stable high traction node $(A_+,C_+)$ we get:
 $\dot{\theta} = -\kappa\sin (\theta)$  with $\kappa = \lambda(1-\varepsilon)kA_+C_+ > 0$. Hence the orientations $\theta = 0$, and $\theta = \pi$ are stable, while $\theta = \pm \pi/2$ are unstable. This confirms the directional alignment of axons observed experimentally (Figure~\ref{fig3}). Incorporating the noise term $\eta(t)$ and performing the standard stochastic averaging we obtain the angular variance \cite{Staii2023, Staii2024}:
\begin{equation}
\bigl\langle \theta^{2} \bigr\rangle = \frac{D_{\theta}}{\kappa}.\tag{14}
\end{equation}
which decreases as traction or mechanical anisotropy increases.

We note that Equation~(13) is dimensionless, because the axonal elongation rate was scaled  with the tubulin polymerization velocity
\(v_0\) (Appendix B). This parameter has been measured in the neurite-growth literature  and has the value \cite{Lowery2009, Franze2010}
$v_0 \approx 30 \mu\mathrm m\,\mathrm h^{-1}$ 

The experimentally measured average speed of the growth cone on micropatterned substrates is $\bar v_{exp} = 10\,\mu \mathrm m\,\mathrm h^{-1}$ \cite{Basso2019, Yurchenko2019, Yurchenko2021, Descoteaux2022}. The measured velocity expressed in these units is therefore:
\begin{equation}
v=\frac{\bar v_{\text{exp}}}{v_0}\approx\frac{10}{30}=0.333 \notag
\end{equation}
which together with Equation~(13) implies $\Lambda \approx 0.17$.

For a growth cone advancing at speed $\bar v_{\text{exp}}$, the mean--square displacement is \cite{Staii2023, Staii2024}:
\begin{equation}
\langle l^{2}(t) \rangle = 2D_{\text{eff}}t,\qquad D_{\text{eff}} = \frac{\bar v_{\text{exp}}D_{\theta}}{2\kappa}.\tag{15}
\end{equation}

\begin{figure*}
    \includegraphics[width=12 cm]{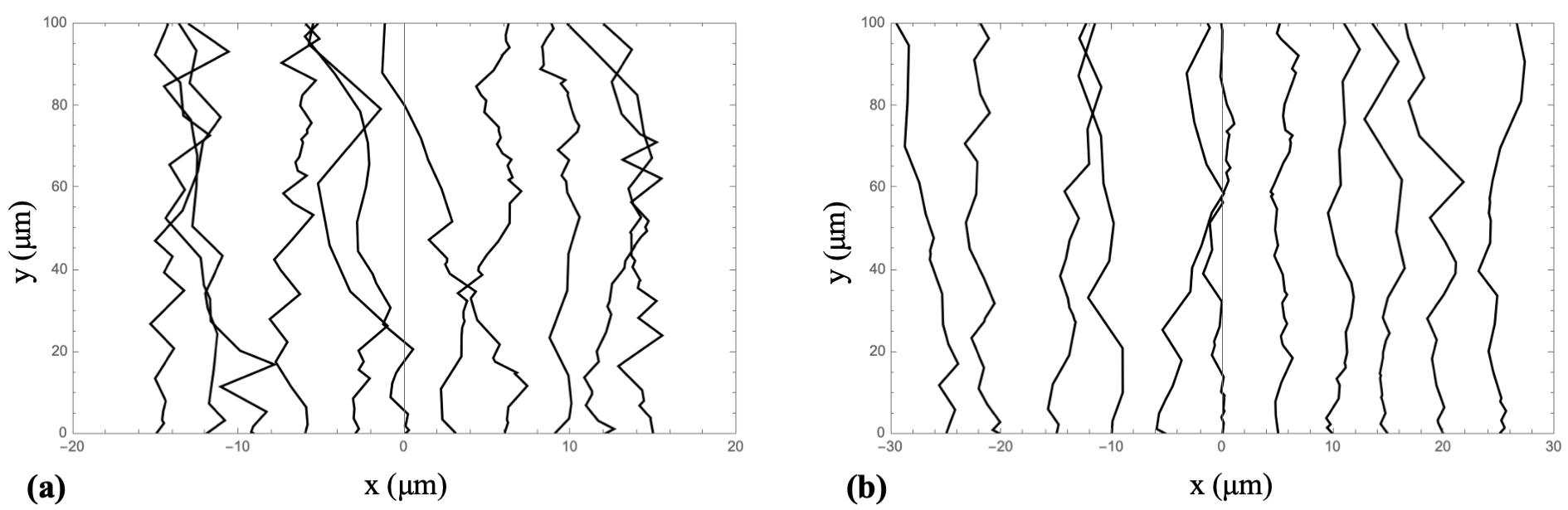}
    \caption{ Simulated axonal growth trajectories for ten neurons cultured on micropatterned PDMS surfaces with spatial periods of $d=3\mu m$ in (\textbf{a}) and $d=5\mu m$ in (\textbf{b}).  Simulations were performed using the theoretical model discussed in the text. The simulated angular alignment, axonal length, and speed distributions are in close agreement with experimental observations.  \label{fig6}}
\end{figure*}

Matching the experimentally measured value for the diffusion coefficient $D_{\text{eff}} = 25\,\mu\mathrm m^{2}\,\mathrm h^{-1}$ implies $D_{\theta}/\kappa = 0.50$. Finally, we use the theoretical model given by Equations~(1)-(11) and the above parameter values inferred from experimental data, to perform numerical simulations of axonal trajectories (Figure~\ref{fig6}). The resulting angular alignment, length, and speed distributions closely match the experimental observations shown in Figure~\ref{fig3}.

\unskip

\section{Discussion}

Despite significant advances in understanding how neurons grow and form functional connections, a comprehensive quantitative framework that accounts for the interactions between the cell and its growth environment in describing axonal dynamics is still missing. Key unanswered questions persist regarding the mechanisms that govern neuronal behavior during axonal migration. A central unresolved question is how coupled positive and negative feedback loops regulate robust axonal growth while enabling adaptive responses to biochemical and mechanical cues. A central question in developmental neurobiology therefore remains: How does the interplay between positive and negative feedback modulate the interactions between intrinsic neuronal behaviors and environmental perturbations to produce reliable morphogenic outcomes?

In the present study we have introduced a unified dynamical--systems framework that integrates four previously disparate aspects of axonal growth: (i) an actin--adhesion traction
switch, (ii) Semaphorin/Slit --mediated lateral inhibition across micropatterns, (iii) soma--to--growth cone tubulin transport, and (iv) orientation dynamics driven by substrate anisotropy. Treating these subsystems
as mutually coupled ordinary differential equations and performing a global
bifurcation analysis reproduces the key phenomena observed for cortical axons
cultured on parallel micropatterns: clutch--like generation of traction forces, constant elongation speed, stable equilibria, and preferential alignment along the grooves. This unification constitutes a conceptual advance over earlier
models that treated each mechanism in isolation \cite{Oliveri2022, Descoteaux2022, Staii2023, Staii2024}. The following paragraphs summarize the key predictions of the mathematical model.

\textit{Emergence of robust high–traction motility}. The reduced two-dimensional subsystem shown in Figure~\ref{fig5}a captures a classical bistable decision process in growth cone dynamics: the system can either collapse into a low-actin, low-adhesion state ($E_0$), or stabilize in a high-actin, high-adhesion state associated with persistent extension ($E_1$). This bistability originates from a positive feedback loop in which increased actin levels promote adhesion assembly, which in turn reinforces actin accumulation, as described by Equations~(1)–(3). The parameters $\alpha$ and $\beta$ govern the turnover rates of actin and adhesions, respectively, and set the timescales that determine the position of the separatrix in phase space. For moderate turnover parameter values ($\alpha = 0.6$, $\beta = 0.5$) the system exhibits two coexisting stable states (attractors), separated by a saddle point, as indicated by the green and red dots in Figure~\ref{fig5}a. Small variations in either turnover rate can remove one equilibrium point through a saddle–node bifurcation, effectively biasing the system toward either collapse or sustained protrusion. For example, Figure~\ref{fig5}b shows that increasing $\alpha$ or $\beta$ removes the high-actin state $E_1$, leaving only the collapsed state at the origin. This result provides a mathematical formulation of a long–standing experimental observation that growth cones behave as bistable
mechanochemical switches, transitioning abruptly from a quiescent state to robust motility once
a minimal level of adhesion signaling is reached \cite{Huber2003, Lowery2009, Franze2010, Kiryushko2004, Kumarasinghe2022}. The explicit criterion derived here could provide a quantitative target for pharmacological manipulation of actin polymerization or
integrin dynamics in regenerative therapies. For example, pharmacological agents that alter actin stability (e.g., taxol) or interfere with myosin-II contractility (e.g. blebbistatin, Y-27632) directly modify $\alpha$ and $\beta$, shifting the separatrix and lowering the probability that a growth cone commits to extension. These theoretical predictions are consistent with our previous experiments, where such treatments markedly impeded growth cone
motility and alignment \cite{Yurchenko2019, Yurchenko2021, Vensi2019, Basso2019, Descoteaux2022}.

\textit{Feedback loop and pattern selection}. Within the mechanochemical signaling component of the axonal‐growth model, the effective control parameter $m = -\,f'(I_0)\,g'(S_0)\,e^{-\gamma F}$ encapsulates the combined sensitivities of inhibitory ($I$) and stimulatory ($S$) signals, along with the influence of the traction force $F$. 
When \(m\) exceeds the critical value $-1$, the uniform signaling state becomes unstable through a supercritical pitchfork bifurcation,
leading to the emergence of two alternating patterns of signaling activity. Since $m$ increases
with traction force $F$, micropatterns with spatial periods that promote focal-adhesion maturation effectively lower the threshold for symmetry breaking and facilitate axonal alignment. At the multicellular level, incorporating traction into the lateral inhibition feedback
loop transforms a standard pitchfork bifurcation into a traction–modulated one. As a result, the critical level of ligand export suppression required for pattern formation is determined by the magnitude of traction force. The model predicts that substrates with specific spatial periods $d$, which enhance traction through improved adhesion maturation, reduce the threshold for transitions between patterned states. This framework thus establishes a mechanistic connection between substrate micro-topography and the patterned alignment of axons observed experimentally. Furthermore, it identifies micropattern periodicity as a tunable design parameter for guiding axonal organization in engineered neural tissues.

\textit{Self-regulating axonal speed and mechanochemical alignment}. Another notable result is the identification of an exact linear solution within the tubulin transport subsystem. Regardless of initial conditions, the system converges to a linearly growing state with a dimensionless velocity given by Equation (13). When converted to physical units, this solution accurately reproduces the observed average growth cone speed of approximately $10 \mu m/hr$. Since this velocity depends solely on the parameter $\Lambda$, which reflects the ratio of tubulin loss to turnover rates, the model predicts that fluctuations in tubulin supply at the growth cone are self-correcting, provided that homeostatic turnover in the soma is preserved. This built-in regulatory mechanism effectively acts as a \textit{speed governor} for axonal elongation, a feature that could play an essential role for coordinated outgrowth during morphogenesis. The orientation Equation~(4) around the micropattern direction shows that the deterministic restoring rate
$\kappa=\lambda(1-\varepsilon)kA^{+}C^{+}$ near a stable high--traction node $(A_+,C_+)$, competes with angular noise
$D_{\theta}$ to determine the variance in orientation $\langle\theta^{2}\rangle=D_{\theta}/\kappa$. The experimentally inferred ratio $D_{\theta}/\kappa\simeq0.5$ agrees with independent measurements of both
traction forces and small‐angle fluctuations \cite{Descoteaux2022, Staii2023, Staii2024}. This supports the conclusion that alignment precision improves with either enhanced actin‐mediated contraction
( increased $A^{+}C^{+}$) or greater anisotropy in substrate stiffness (larger $(1-\varepsilon)$). From an applied perspective, the model offers explicit design guidelines for engineering biomaterial micropatterns: increasing ridge depth or enhancing the contrast in Young’s modulus should significantly reduce transverse diffusion (i.e. effective $D_{\text{eff}}$) without affecting the rate of axonal extension.

\textit{Future directions}. Together, these results advance our understanding of axon guidance on three
fronts.  First, they demonstrate that complex neurite behaviors can be
captured by low‐dimensional dynamics once the correct
\emph{mechanochemical couplings} are included, endorsing dynamical--systems
theory as a unifying framework for investigating neuronal growth.  Second, the model yields
closed‐form, testable expressions for experimentally accessible observables
(speed, bundle spacing, diffusion coefficient), enabling
\emph{parameter‐free} predictions that can be falsified or refined by future
traction‐force microscopy and live‐imaging experiments.  Third, the framework
is readily extensible: adding chemotropic gradients, growth‐cone steering
torques, or extracellular‐matrix degradation terms requires only minor
modifications to the existing set of ordinary differential equations.

From a translational standpoint, identifying a minimal set of dimensionless
groups that govern alignment and speed paves the way for rational design of
micropatterned scaffolds in neural repair.  Manipulating groove anisotropy or
modulating adhesion turnover could, for instance, optimize both the direction
and rate of axonal regrowth after spinal‐cord injury.

\section{Conclusions}

In this study, we present an integrated biophysical model of neuronal growth on micropatterned substrates that incorporates molecular transport, cell–substrate mechanics, and cell–cell signaling. This unified framework enables accurate interpretation and prediction of axonal dynamics and network formation within engineered microenvironments. By reproducing experimental observations with minimal adjustable parameters, the model offers valuable insights for both fundamental neuroscience research and the development of bioinspired therapeutic strategies.


\vspace{6pt} 



\section*{funding}
This research was funded by Tufts University Faculty Research Award and Tufts Summer Faculty Fellowship Award.

\section*{acknowledgments}
We thank Joao Marcos Vensi Basso, Ilya Yurchenko, and Marc Descoteaux for their participation during the initial stages of this project.

\section*{conflicts of interest}
The authors declare no conflicts of interest. The funders had no role in the design of the study; in the collection, analyses, or interpretation of data; in the writing of the manuscript; or in the decision to publish the results.





\appendix
 \section[\appendixname~\thesection]{Steady states of the biomechanical equations and pitchfork bifurcation} 

The coupled actin ($A$) and adhesion complex ($C$) densities follow Equations (1)-(3) in the main text.
Setting $\dot A=\dot I=0$ and dividing by $A>0$ 
leads to the quadratic equation: $\alpha \beta A^{2}-A+\alpha \beta =0$
This equation has two positive roots when $0<\alpha  \beta<\frac12$, a
double root at $\alpha  \beta =\frac12$, and no positive root when $\alpha  \beta>\frac12$.
Hence, a \emph{saddle-node bifurcation} occurs at
\begin{equation}
\alpha  \beta=\tfrac12 \tag{A1}.
\end{equation}
Explicitly,
\begin{equation}
A_{\pm}=\frac{1\pm\sqrt{1-4(\alpha  \beta)^{2}}}{2\alpha  \beta},\qquad 
I_{\pm}=\alpha A_{\pm},\qquad 0<\gamma<\tfrac12 . \tag{A2}
\end{equation}

The Jacobian of the biomechanical system is:
\begin{equation}
  J=
\begin{pmatrix}
-\alpha & 1\\[2pt]
\displaystyle\frac{2A}{(1+A^{2})^{2}} & -\beta
\end{pmatrix}. \tag{A3}  
\end{equation}

Its trace is negative ($-\alpha-\beta<0$), so fixed points are either stable nodes
or saddles according to $\det J$:
\[
\det J = \alpha\beta - \frac{2A}{(1+A^{2})^{2}}.
\]
At $(A,I)=(0,0)$ we have $\det J = \alpha\beta>0$ $\Rightarrow$ stable node.  
For $\alpha\beta<\frac12$, the larger positive root $(A_{+},I_{+})$ is also a stable
node, whereas the smaller root $(A_{-},I_{-})$ is a saddle
($\det J<0$ because $A_{-}<1$).

For the periodic array of micropatterns we decompose near-neighbor biochemical inhibition into Fourier modes. Linearizing Equations (5)-(8) about the homogeneous fixed point and Fourier transforming $\{I_i,S_i\}\propto e^{ikx}$ yields the Jacobian:
\begin{equation}
J(k) = \begin{pmatrix}-1 & f'\cos k\\ -\nu g' e^{-\gamma F} & -\nu e^{-\gamma F}\end{pmatrix} \tag{A4}    
\end{equation}
whose eigenvalues are:
\begin{align}
\Lambda_{\pm}(k) &= \tfrac{1}{2} \bigl[ -(1 + \nu e^{-\gamma F}) \notag \\
&\quad \pm \sqrt{(1 - \nu e^{-\gamma F})^{2} 
- 4 \nu e^{-\gamma F} f' g' \cos k} \bigr] \tag{A5}
\end{align}

Setting the spatial wavenumber $k = \pi$, and the control parameter $m \;\equiv\; -\,f'(I_0)\,g'(S_0)\,e^{-\gamma F}=-1$ defines the threshold for the pitchfork bifurcation in lateral inhibition discussed in the main text.

\section{Tubulin transport}
The one--dimensional convection--diffusion equation for tubulin concentration
$c(x,t)$ along an axon of instantaneous length $L(t)$ is \cite{Oliveri2022}:
\begin{equation}
\partial_t c + \partial_x J = S, \qquad
J = D\,\partial_x c + a c, \qquad S = -\lambda c. \tag{A6}
\end{equation}
\noindent
Here $J$ denotes the total tubulin flux from the soma, comprising passive diffusion with coefficient $D$ and an active transport component mediated by molecular motors, given by $ac$. $S$ accounts for any processes that add to or remove material from the system, acting as a general source or sink. Axonal growth depends on the balance between tubulin assembly and disassembly, which in turn regulates growth cone advance through the condition:

\begin{equation}
\frac{1}{e}\,\dot{L}=k_+\,c_1(t)-k_-,
\tag{A7}\end{equation}
where $c_1(t)$ is the tubulin concentration in the axon, $e$ is the elongation per incorporated dimer, and $k_+$, $k_-$ are the rate constants. 

In addition, tubulin is supplied by the soma with rate:

\begin{equation}
\frac{dc_0}{dt}=S-J\cdot \frac{V}{A},
\tag{A8}
\end{equation}
where $A$ is the axonal cross-sectional area and $V$ is the effective volume involved in tubulin transport.

Following Oliveri\,\&\,Goriely  \cite{Oliveri2022} we scale time by $A/D$, length by $e(SA/D)$, and concentration by $SA/VD$ to obtain the dimensionless Equations (9)-(11) in the main text:
\begin{align}
\dot{c}_0 &= 1 - \alpha_t\frac{c_1 - c_0}{\ell}, \tag{9}\\
\dot{c}_1 &= -\gamma_t c_1 + \beta_t + \alpha_t\frac{c_1 - c_0}{\ell}, \tag{10}\\
\dot{\ell} &= \gamma_t c_1 - \beta_t,\tag{11}
\end{align}

The one–parameter family of solutions for this system has the form \cite{Oliveri2022}:
\begin{align}
c_1(t) &= c_{00} + c_{01}t, \tag{A9} \\
c_{01} &= 1-\frac{\sqrt{\Lambda(\Lambda+4)}-\Lambda}{2}, \tag{A10}\\
v &= \frac{\sqrt{\Lambda(\Lambda+4)}-\Lambda}{2}, \tag{A11}
\end{align}
where $\Lambda=\beta_t/\gamma_t$ (dimensionless disassembly/assembly ratio). Along these trajectories the axon elongates at constant speed $v$ while the soma concentration remains time–independent. From $\dot\ell=\gamma_t c_1-\beta_t$ we see that initially:
\begin{itemize}
  \item $c_1(0)<\beta_t/\gamma_t\;(=\Lambda)\;\Rightarrow\;\dot\ell(0)<0$ (axon shortens),
  \item $c_1(0)>\Lambda\;\Rightarrow\;\dot\ell(0)>0$ (axon elongates).
\end{itemize}
However, Equations (9)-(11) imply that $\dot c_1$ and $\dot\ell$ always have opposite signs when $|c_1-c_0|$ is small, so $c_1$ is driven towards $\Lambda$ while $\dot\ell$ changes sign only once. The analytical solution confirms this: even when a transient collapse occurs the system subsequently switches to linear growth once $c_1$ crosses $\Lambda$ (Figure~\ref{fig4}a). Consequently, the long–time dynamics of Equations (9)-(11) are universally governed by the linear–growth regime regardless of initial conditions.







\bibliographystyle{apsrev4-2}
\bibliography{NewCitation}

\end{document}